\documentclass[a4paper,cite,11pt,psnfss]{article}

\usepackage[latin1]{inputenc}
\usepackage{comment}
\usepackage{ulem}
\usepackage{adjustbox}
\usepackage{appendix}
\usepackage{framed,epsfig,pgfplots,amsmath}
\usepackage{tikz}
\usepackage{tikz-feynhand}
\usetikzlibrary{snakes,arrows,shapes,positioning,automata,backgrounds,calc,er,patterns}
\usepackage{caption}
\usepackage{subcaption}
\usepackage{cite}
\usepackage[colorlinks,citecolor=blue]{hyperref}
\usepackage{mathtools, amsmath, amsthm, amssymb,amsfonts}
\usepackage{makecell}
\usepackage{bm}
\usepackage{url}
\usepackage{hhline}
\usepackage{multirow}

\usepackage[usenames,dvipsnames,tree]{pstricks}

\definecolor{boxcolor}{RGB}{235,245,255}

\usepackage{graphicx,color,pst-plot}

\usepackage{latexsym,amsmath,amsfonts,amssymb,amstext,mathrsfs,upgreek}

\allowdisplaybreaks[1]

\begin{document}

\begin{titlepage}
\begin{flushright}
PSI-PR-24-05, ZU-TH 05/24
\end{flushright}
\begin{flushright}
\end{flushright}

\vfill

\begin{center}

{\Large\bf Geometrical methods for the analytic evaluation}
\medskip

{\Large\bf of multiple Mellin-Barnes integrals$^{\star}$}


\vfill

{\bf Sumit Banik$^{a,b\dagger}$ and Samuel Friot$^{c,d\ddagger}$}\\[1cm]
{$^a$ Physik-Institut, Universitat Zurich,\\ Winterthurerstrasse 190, CH 8057 Zurich, Switzerland}\\[0.5cm]
{$^b$ Paul Scherrer Institut, CH 5232 Villigen PSI, Switzerland}\\[0.5cm]
{$^c$ Universit\'e Paris-Saclay, CNRS/IN2P3, IJCLab, 91405 Orsay, France } \\[0.5cm]
{$^d$ Univ Lyon, Univ Claude Bernard Lyon 1, CNRS/IN2P3, \\
 IP2I Lyon, UMR 5822, F-69622, Villeurbanne, France}\\[0.5cm]
\end{center}
\vfill

\begin{abstract} 
Two recently developed techniques of analytic evaluation of multifold Mellin-Barnes (MB) integrals are presented. Both approaches rest on the definition of geometrical objets conveniently associated with the MB integrands, which can then be used along with multivariate residues analysis to derive series representations of the MB integrals. The first method is based on introducing conic hulls and considering specific intersections of the latter, while the second one rests on point configurations and their regular triangulations. After a brief description of both methods, which have been automatized in the \texttt{MBConicHulls.wl} \textit{Mathematica} package, we review some of their applications. In particular, we show how the conic hulls method was used to obtain the first analytic calculation of complicated Feynman integrals, such as the massless off-shell conformal hexagon and double-box. We then show that the triangulation method is even more efficient, as it allows one to compute these nontrivial objects and harder ones in a much faster way.

\end{abstract}

\vspace{1cm}

\small{$\dagger$ sumit.banik@psi.ch}

\small{$\ddagger$ samuel.friot@universite-paris-saclay.fr}

\vspace{1cm}

$^{\star}$ Presented by S. Friot at the XLV International Conference of Theoretical Physics ``Matter to the Deepest", Warsaw, Poland, 17-22 September, 2023 (extended version of the contribution to the proceedings published in Acta Physica Polonica B).
\end{titlepage}

\section{Introduction}

This talk contribution is based on \cite{Ananthanarayan:2020fhl} and subsequent works \cite{Banik:2022bmk, Banik:2023rrz, Ananthanarayan:2020ncn, Ananthanarayan:2020xpd, Friot:2022dme}, where the focus is on methods of analytic calculation of multiple Mellin-Barnes (MB) integrals, their automatization and some of their possible applications. Such integrals are useful in many domains of physics and mathematics such as quantum field theory \cite{Smirnov:2012gma, Dubovyk:2022obc}, electromagnetic waves in turbulence \cite{Sasiela}, asymptotics \cite{Paris}, hypergeometric functions \cite{Marichev}, etc. The need for a systematic method of analytic computation of MB integrals was recently emphasized for instance in \cite{Loebbert:2019vcj}, in the quantum field theory context, more specifically for the computation of Feynman integrals. Indeed, Feynman integrals are fundamental objects for scattering amplitudes in  particle physics and one possible way to evaluate them is through their MB integral representations. 

The \texttt{MBConicHulls.wl} \textit{Mathematica} package, whose first version was made public in \cite{Ananthanarayan:2020fhl} and that we further developed in  \cite{Banik:2022bmk, Banik:2023rrz}, is an efficient tool for the analytic computation of MB integrals which rests on two original geometrical methods that we will present in this talk.
In fact, prior to this work, there has been one attempt to build a \textit{Mathematica} package dedicated to this task and called \texttt{MBsums.m}, which was published in the proceedings of the MTTD15 conference \cite{Ochman:2015fho}. In the approach developed by these authors, the MB integrations are performed sequentially which, depending on their order, can in principle provide at the end different series representations of the MB integral. However, it does not seem that the results given by the \texttt{MBsums.m} package are always optimal, since when applied to the MB representations of complicated Feynman integrals, for instance, an explosion of the number of terms renders the obtained analytical expressions difficult to use. One may thus wonder if the iterative integration procedure is the right approach to follow. Moreover, the \texttt{MBsums.m} package, in its present version, is restricted to the evaluation of Mellin-Barnes integrals with straight contours of integration. Therefore, in addition to the above mentioned motivations coming from \cite{Loebbert:2019vcj}, these two drawbacks of \texttt{MBsums.m} pushed us to revisit this problem and, inspired by the works of some mathematicians \cite{Tsikh2, Passare:1996db, Tsikh3}, our study gave birth to two non-iterative methods which can be used for the evaluation of multiple MB integrals with straight or non-straight contours, and which dramatically reduce the number of terms in the final expressions. These two new methods rest on the introduction, during the intermediate calculational steps, of geometrical objects associated with the MB integrand, and give the final results in terms of (multivariable) series representations of the hypergeometric type. The first method uses conic hulls, specific intersection of which yielding different series representations of the MB integral \cite{Ananthanarayan:2020fhl}, while the second method is based on triangulations of some configurations of points, each triangulation being in one-to-one correspondence with one series representation \cite{Banik:2023rrz}. These two equivalent geometrical intermediate steps are followed by a multivariate residues analysis and, as already said above, the whole corresponding calculational procedures have been automated in the \texttt{MBConicHulls.wl} \textit{Mathematica} package \cite{MBConicHullsGit}. Although they give exactly the same final results, the triangulation method is much faster than the conic hulls method when used for the computation of complicated multifold MB integrals.

The outline of this contribution is as follows. In a first introductory part, we will recall what MB integrals are and give a short motivation for their study. In particular, their historical link to the hypergeometric functions theory where they can be used, among others, to study their transformation theory, will be briefly presented as well as their use as a tool for the computation of Feynman integrals. One will then switch to the main core of the paper which concerns the analytic evaluation of MB integrals. After summarizing  known facts, we will present our two methods. Then, a brief description of some recent applications of the latter will be followed by our conclusions.

\section{What is a Mellin-Barnes integral?\label{Section1}}

\begin{figure}
        \centering
        \includegraphics[scale=0.45]{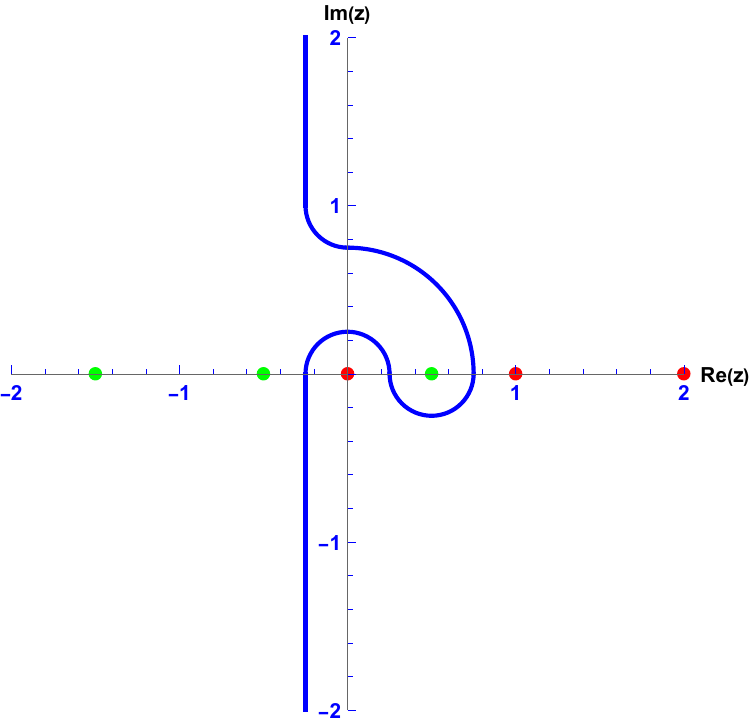}
       \caption{\label{contour}The integration contour of Eq. (\ref{1foldEx}) (in blue). It separates the poles of $\Gamma(-z)$ (in red) from the poles of $\Gamma(-1/2+z)$ (in green).}
    \end{figure}

Let us first present this particular type of integrals in the simplest situation, \textit{i.e.} the one-fold case. The general expression of such MB integrals reads
\begin{equation}
I_1=\int\limits_{-i \infty}^{+i \infty}  \frac{d z}{2 \pi i} \, x^z\, \,  \frac{\prod\limits_{i=1}^{k} \Gamma^{a_i}( e_i z+g_i)}{\prod\limits_{j=1}^{l} \Gamma^{b_j}(f_j z+h_j)}    \label{1foldMB}
\end{equation}
where $a_i , b_j, k$ and $l$ are positive integers and  $e_i, f_j$ are real numbers while $g_i, h_j$ can be complex. When not specified, as in Eq. (\ref{1foldMB}), it is implicit that the contour of integration is such that, for each of the gamma functions of the numerator of the MB integrand, it does not separate the corresponding set of poles in different subsets. This implies that for some MB integrals, the contour cannot be a straight line parallel to the imaginary axis, as in the following particular example
  \begin{align}
   I=  \int\limits_{-i \infty}^{+i \infty}  \frac{d z}{2 \pi i} (-x)^{z}\,\Gamma(-z)\Gamma(-1/2+z) \label{1foldEx}
  \end{align}
  where the integrand has no denominator and has poles at $z= 0 ,1 ,2 , \cdots$ (due to $\Gamma(-z)$) and at $z= 1/2 , -1/2 ,-3/2 , \cdots$ (from $\Gamma(-1/2+z)$). In Eq. (\ref{1foldEx}) the integration contour is not specified, which means from the rule above that it must follow some path in the complex $z$-plane such as the one shown in Fig. \ref{contour}. 

Now that we haved fixed ideas, one can consider the general $N$-fold case, where MB integrals read
  \begin{align}
  I_N=\int\limits_{-i \infty}^{+i \infty} \frac{ d z_1}{2 \pi i} \cdots \int\limits_{-i \infty}^{+i \infty}\frac{ d z_N}{2 \pi i}\,\, \,  x^{z_1}_{1} \cdots x^{z_N}_{N} \, \, \, \frac{\prod\limits_{i=1}^{k} \Gamma^{a_i}({\bf e}_i\cdot{\bf \mathbf{z}}+g_i)}{\prod\limits_{j=1}^{l}\Gamma^{b_j}({\bf f}_j\cdot{\bf z}+h_j)}\label{NfoldMB}
  \end{align}
  where ${\bf e}_i$, ${\bf f}_j$ are $N$-dimensional coefficient vectors and ${\bf z}=(z_1, ... , z_N)$ and the main question addressed in this contribution is how such integrals can be computed analytically. 
    
If one considers Eq.(\ref{1foldEx}) there is a simple way to do this. One can close the contour of integration following a half circle at infinity either to the left, or to the right. Indeed, by means of a constraint on the $x$ scale and due to the sufficiently decreasing behavior of the gamma functions on this half circle, adding it to the original MB integral will not change its value. Now, once the contour is closed, one can use Cauchy's theorem to compute the integral by summing the residues at the poles that are located inside the contour. This way, closing the contour to the right, one obtains the following series representation of the integral
     \begin{align}
          \sum_{n=0}^{\infty} \Gamma(-1/2+n) \frac{x^{n}}{n!} \hspace{1cm} |x|<1
      \end{align}
  while closing to the left gives 
      \begin{align}
        (-x)^{1/2}  \sum_{n=0}^{\infty} \Gamma(-1/2+n) \frac{x^{-n}}{n!} \hspace{1cm} |x|>1
      \end{align}
      In this simple case, both series can be resummed, yielding the same final result $-2\sqrt{\pi}\sqrt{1-x}$ which in fact is a particular case of the well-known MB representation used in QFT
\begin{align}
      \frac{1}{(A+B)^{\alpha}}=\frac{1}{\Gamma(\alpha)}\int\limits_{-i \infty}^{+i \infty}  \frac{d z}{2 \pi i} \Gamma(-z) \Gamma(\alpha+z) A^{-\alpha-z} B^{z}\label{MBqft}
  \end{align}
In the multifold case, one can in principle follow the same approach and, applying Cauchy's theorem iteratively, obtain several multiple series representations (it is however difficult in general to perform the resummation of these series in terms of standard functions as in the simple example above). In the case of straight contours, this is the strategy that has been adopted and automated in \cite{Ochman:2015fho}. As mentioned in the introduction of this contribution we have developed alternative methods which are non-iterative and valid for straight or non-straight contours. However, before describing them, we propose a short digression into the mathematical context in which MB integrals were historically recognized as a useful tool\footnote{Details about the birth of MB integrals can be found in \cite{Pincherle}.}, namely the theory of hypergeometric functions, and how later, in physics, they were used for the computation of Feynman integrals in quantum field theory.

\section{MB integrals, hypergeometric functions and Feynman integrals}
\subsection{MB integrals and hypergeometric functions}
\begin{figure}[h]
       \centering
       \includegraphics[width=8cm,angle =0.5 ]{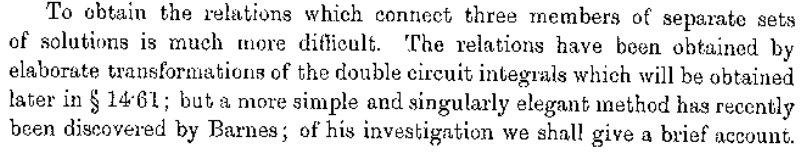}
       \caption{Whittaker and Watson, \textit{``A course of Modern Analysis"} 1927, p.286.  \label{WW}}
   \end{figure}

It was Barnes, in 1908, who obtained the MB representation of the celebrated Gauss ${}_2F_1$ hypergeometric function \cite{Barnes}
\begin{align}	
	{}_2F_1(a,b;c;z)&=\frac{\Gamma(c)}{\Gamma(a)\Gamma(b)}\int\limits_{-i \infty}^{+i \infty}  \frac{ds}{2i\pi}(-z)^s\Gamma(-s)
	\frac{\Gamma(a+s)\Gamma(b+s)}{\Gamma(c+s)}
	\label{MB2F1}
\end{align}
where the contour of integration separates the poles of $\Gamma(-s)$ from those of  $\Gamma(a+s)$ and  $\Gamma(b+s)$. Closing the contour to the right as performed in the simple example of the previous section, one obtains the well-known series representation
\begin{align}
	{}_2F_1(a,b;c;z)&=\frac{\Gamma(c)}{\Gamma(a)\Gamma(b)}\sum_{n=0}^\infty\frac{\Gamma(a+n)\Gamma(b+n)}{\Gamma(c+n)}\frac{z^n}{n!}
\end{align}
As noted by Whittaker and Watson in their famous book \cite{W&W}, that we quote in Fig. \ref{WW}, the interest in the MB representation given in Eq.(\ref{MB2F1}) is that it analytically continues this series and, by closing the contour to the left, it provides the following transformation formula 
\begin{align}
	{}_2F_1(a,b;c;z)=\frac{\Gamma(c)\Gamma(b-a)}{\Gamma(c-a)\Gamma(b)}&(-z)^{-a}{}_2F_1(a,a-c+1;a-b+1;1/z)\nonumber \\
	+\frac{\Gamma(c)\Gamma(a-b)}{\Gamma(c-b)\Gamma(a)}&(-z)^{-b}{}_2F_1(b,b-c+1;b-a+1;1/z)\label{transf2F1}
\end{align}
which is probably the best known of the relations that link three solutions of the  differential equation satisfied by the Gauss ${}_2F_1$ hypergeometric function.
   
  This elegant approach of finding the transformations of ${}_2F_1$ did not escape to Appell, who introduced his four functions $F_1$, $F_2$, $F_3$ and $F_4$ extending Gauss' ones at the two variable level, as one can read in the book that he published with Kamp\'e de F\'eriet \cite{KdF} where they give their MB representations and use the latter to study their transformation theory. As an example, we reproduce here the MB representation of the Appell $F_1$ function, which reads
\begin{align}
        F_1(a,b_1,b_2;c;x,y)&= \frac{\Gamma(c)}{\Gamma(a)\Gamma(b_1)\Gamma(b_2)}\int\limits_{-i \infty}^{+i \infty}  \frac{d z_{1}}{2 \pi i}  \int\limits_{-i \infty}^{+i \infty}  \frac{d z_{2}}{2 \pi i}(-x)^{z_{1}} (-y)^{z_{2}} \Gamma(-z_1)\Gamma(-z_2)\nonumber \\ &
\times   \frac{\Gamma\left(a+z_{1}+z_{2}\right) \Gamma\left(b_1+z_{1}\right) \Gamma\left(b_2+z_{2}\right)}{\Gamma\left(c+z_{1}+z_{2}\right)}  
    \end{align}
Using the \texttt{MBConicHulls.wl} package, it is a straightforward exercise to find the transformation formulas of $F_1$, $F_2$, $F_3$ and $F_4$ by a direct evaluation of their MB representations. Doing so, one obtains four linear transformations for $F_1$ and $F_2$, three for $F_3$ and two for $F_2$, which can be seen as the generalizations of Eq.(\ref{transf2F1}) and whose expressions can be found in \cite{Srivastava}. As described in the latter reference, many other extensions of the Gauss ${}_2F_1$ exist. The methods that we have developed and the \texttt{MBConicHulls.wl} package can naturally be used to study their transformation theory, as shown in \cite{Ananthanarayan:2020xpd, Friot:2022dme} where a focus has been put on the Srivastava $H_c$ three variables extension
\begin{align}
	H_C(a,b,c;d;x,y,z)=\frac{\Gamma(d)}{\Gamma(a)\Gamma(b)\Gamma(c)}\sum_{m,n,p=0}^\infty\frac{\Gamma(a+m+n)\Gamma(b+m+p)\Gamma(c+n+p)}{\Gamma(d+m+n+p)}\frac{x^m}{m!}\frac{y^n}{n!}\frac{z^p}{p!}
\end{align}
whose MB representation is
\begin{align}
	H_C(a,b,c;d;x,y,z)&=\frac{\Gamma(d)}{\Gamma(a)\Gamma(b)\Gamma(c)}\int\limits_{-i \infty}^{+i \infty}\frac{dz_1}{2i\pi}\int\limits_{-i \infty}^{+i \infty}\frac{dz_2}{2i\pi}\int\limits_{-i \infty}^{+i \infty}\frac{dz_3}{2i\pi}(-x)^{z_1}(-y)^{z_2}(-z)^{z_3}\nonumber\\
	&\times \Gamma(-z_1)\Gamma(-z_2)\Gamma(-z_3)\frac{\Gamma(a+z_1+z_2)\Gamma(b+z_1+z_3)\Gamma(c+z_2+z_3)}{\Gamma(d+z_1+z_2+z_3)}\label{MBHC}
\end{align}
This three variables hypergeometric function is linked to the massless conformal triangle Feynman integral \cite{Loebbert:2020hxk} and in \cite{Ananthanarayan:2020xpd}, the thirteen linear transformations of $H_C$ that can be derived from a direct calculation of Eq.(\ref{MBHC}) were given as an illustration of the conic hulls computational method, while in \cite{Friot:2022dme}, hundreds of other (and in general new) transformations are obtained, following the method employed by Barnes and nicely detailed in \cite{W&W}. As an example, we give below one of the simplest of these transformations (which was already known \cite{Srivastava}): 
\begin{align}
  H_C(a,b,c;d;x,y,z) 
  &= \frac{\Gamma(d)\Gamma(c-b)}{\Gamma(d-b)\Gamma(c)}(-z)^{-b}G_C(b,a,b-d+1;b-c+1;x/z,1/z,y) \nonumber\\ 
  & + \frac{\Gamma(d)\Gamma(b-c)}{\Gamma(d-c)\Gamma(b)}(-z)^{-c}G_C(c,a,c-d+1;c-b+1;y/z,1/z,x)
\end{align}
where 
\begin{align}
G_C(a,b,c;d;x,y,z) =\sum\limits_{m,n,p=0}^{\infty}\frac{(a)_{m+n}(b)_{m+p}(c)_{n-p}}{(d)_{m+n-p}} \frac{x^{m}y^{n}z^{p}}{m! \, n! \, p!}
\end{align}
is another Srivastava series \cite{Srivastava}.

Very few of the multivariable extensions of ${}_2F_1$ have been studied in this way, this first analysis thus shows that the \texttt{MBConicHulls.wl} package is a promising tool for the exploration of the transformation theory of multivariable hypergeometric functions.
\subsection{MB integrals and Feynman integrals}
As noted in \cite{Regge} almost sixty years ago, there is a close link between hypergeometric functions and Feynman integrals. It is thus natural to expect that MB integrals can play a role in the evaluation of Feynman integrals and this has been explicitly shown in 1975 by Usyukina \cite{Usyukina:1975yg}. Indeed, applying Eq.(\ref{MBqft}) to the propagators, or to the parametric representations of Feynman integrals, one can obtain MB representations of the latter. This approach has been widely used during the nineties and beginning of the 2000's (see \cite{Smirnov:2012gma} and references therein) until its automatization in the \texttt{AMBRE} \textit{Mathematica} package \cite{Gluza:2007rt, Ambre} (see also \cite{Belitsky:2022gba} for a recent alternative package).

As an illustration we show below two different examples of MB representations of Feynman integrals (written up to some overall factors): the 3-fold MB representation of the off-shell one loop 3-point scalar Feynman integral with three equal masses \cite{Davydychev:1990jt}
\begin{align}
	&\int\limits_{-i \infty}^{+i \infty}\frac{dz_1}{2i\pi}\int\limits_{-i \infty}^{+i \infty}\frac{dz_2}{2i\pi}\int\limits_{-i \infty}^{+i \infty}\frac{dz_3}{2i\pi}
	(-X_1)^{z_1}(-Y_1)^{z_2}(-Z_1)^{z_3}\Gamma(-z_1)\Gamma(-z_2)\Gamma(-z_3)\nonumber\\
	&\times \Gamma(\nu_1+z_1+z_2)\Gamma(\nu_2+z_2+z_3)\Gamma(\nu_3+z_1+z_3)
	\frac{\Gamma(\nu_1+\nu_2+\nu_3-n/2+z_1+z_2+z_3)}{\Gamma(\nu_1+\nu_2+\nu_3+2z_1+2z_2+2z_3)}
\end{align}
and the 4-fold MB representation of the off-shell one loop scalar massless pentagon \cite{DelDuca:2009ac}
\begin{align}
&\int\limits_{-i \infty}^{+i \infty}\frac{dz_1}{2i\pi}\int\limits_{-i \infty}^{+i \infty}\frac{dz_2}{2i\pi}\int\limits_{-i \infty}^{+i \infty}\frac{dz_3}{2i\pi}
	\int\limits_{-i \infty}^{+i \infty}\frac{dz_4}{2i\pi}(-X_2)^{z_1}(-Y_2)^{z_2}(-Z_2)^{z_3}(-T)^{z_4}\Gamma(-z_1)\Gamma(-z_2)\Gamma(-z_3)\Gamma(-z_4)\nonumber\\
	&\times \Gamma(\nu_5+z_1+z_2)\Gamma(n/2-N_\nu+\nu_1-z_1-z_2-z_3)\Gamma(\nu_2+z_2+z_3)\nonumber\\
	&\times\Gamma(n/2-N_\nu+\nu_3-z_2-z_3-z_4)\Gamma(\nu_4+z_3+z_4)\Gamma(-n/2+N_\nu+\nu_3+z_1+z_2+z_3+z_4)
\end{align}
where the $\nu_i$ $(i=1,2,3,4)$ are the powers of propagators with $N_\nu=\sum_{i=1}^4\nu_i$, $n$ is the space-time dimension, and we refer the reader to \cite{Davydychev:1990jt, DelDuca:2009ac} for the precise definitions of the scales $X_i, Y_i, Z_i$ $(i=1,2)$ and $T$ which are combinations of the squared external momentum (and the mass in the 3-point case). 

For more details about MB integrals and their applications in quantum field theory, we refer the reader to the recent book \cite{Dubovyk:2022obc}. 
Let us now switch to the heart of the paper, which concerns the analytic computation of MB integrals and, in particular, the presentation of the new geometrical methods that we have developed and implemented in the \texttt{MBConicHulls.wl} package.

\section{Evaluating multiple MB integrals analytically}

The analytic evaluation of one-fold MB integrals is mathematically well understood and fully described in books such as \cite{Marichev, Paris}. The multifold case is much less tamed. Several theoretical physicists worked a lot on the elaboration of automatized tools necessary for the computation of MB representations of Feynman integrals (see for instance \cite{MBtools}) and, in this context, as mentioned previously in this talk, the \texttt{MBsums.m} \textit{Mathematica} package of \cite{Ochman:2015fho}, which can be used for an iterative evaluation of multiple MB integrals with straight contours, was built a few years ago.

Nearly in parallel to these investigations, some mathematicians developed a non-iterative and rigorous approach based on the theory of mutidimensional residues \cite{Tsikh1}, in a series of papers \cite{Tsikh2, Passare:1996db, Tsikh3}. However, to our knowledge, they did not explicitly work out the formulas necessary for the evaluation of $N$-fold MB integrals when $N>2$, even in the non-logarithmic (or non-resonant) case\footnote{The $N=2$ logarithmic case was considered in \cite{Friot:2011ic} (for MB integrals with straight contours).}. Only the $N=2$ case is explicitly described in these papers. Inspired by these works and motivated to fill this gap, we developed a first alternative and efficient method \cite{Ananthanarayan:2020fhl} based on conic hulls which, although not established with the same level of rigor as in \cite{Tsikh2, Passare:1996db, Tsikh3}, has been tested on a considerable number of examples, either analytically when results were available for comparison, or numerically. We are now going to describe this method in more details but for this,  we have to first recall a few general facts about MB integrals.

There are two main types of MB integrals :
\begin{itemize}
     \item The degenerate\footnote{There is in fact another, slightly different, degenerate case, see \cite{Tsikh3}.} (also called ``balanced") type, where ${\bf \Delta}\doteq\sum a_i  {\bf e_i} - \sum b_j {\bf f_j}={\bf 0}$. In this case, several series representations of the corresponding MB integral coexist which are in general analytic continuations of one another. MB representations of Feynman integrals belong to this class.
 
     \item The non-degenerate type, which satisfies ${\bf\Delta} \neq {\bf 0}$. In the latter case, one or more convergent series converge for all values of the scales. Additionally, asymptotic series also arise. 
 \end{itemize}
 MB integrals can be further classified regarding the singularity structure of their integrand:
  \begin{itemize}
      \item  The non-resonant (also called ``generic") case: Here, the number of singular hyper-planes intersecting at any pole is equal to the number of folds of the MB integral. 
      
      \item The resonant (also called ``logarithmic") case, where the number of singular hyper-planes intersecting at some or all poles is greater than the number of folds of the MB integral, giving birth to logarithmic contributions in the series representations.
  \end{itemize} 
 The conic hulls/triangulation methods described below can be used to compute MB integrals falling into any of these classes.
 
\subsection{Brief overview of the conic hulls method (non-resonant case)}

The main steps of the conic hulls method, in the non-resonant situation, can be summarized as follows (we refer the reader to \cite{Ananthanarayan:2020fhl} for more details)
\begin{itemize}
\item \textbf{Step 1:} Find all possible $N$-combinations  (where $N$ stands for the number of folds of the MB integral) of the numerator gamma functions and retain non-singular ones.

\item \textbf{Step 2:} Associate a series (building block) with each combination (obtained from a residue calculation at the poles of the corresponding gamma functions).

\item \textbf{Step 3:} Construct a conic hull for each combination/series.

\item \textbf{Step 4:} The largest intersecting subsets of conic hulls give series representations of the MB integral which simply are the sums of the corresponding building blocks.

\item \textbf{Step 5:} The intersecting region gives the master conic hull. The latter is very useful to derive a master series, which can considerably simplify the convergence analysis of the series representation.
\end{itemize}

These computational steps have been automatized in a  \textit{Mathematica} package called \texttt{MBConicHulls.wl} \cite{Ananthanarayan:2020fhl, MBConicHullsGit}.

The resonant case is more complicated because it rests on a non-trivial multivariate residues analysis: the series representations cannot be simply reduced to sums of building blocks as in the non-resonant case (see \cite{Ananthanarayan:2020fhl} for more details). In order to handle the resonant case with the \texttt{MBConicHulls.wl} package, we internally used the \texttt{MultivariateResidues.m} \textit{Mathematica} package \cite{Larsen:2017aqb}. 

\begin{figure}[h]
       \centering
       \includegraphics[width=4cm]{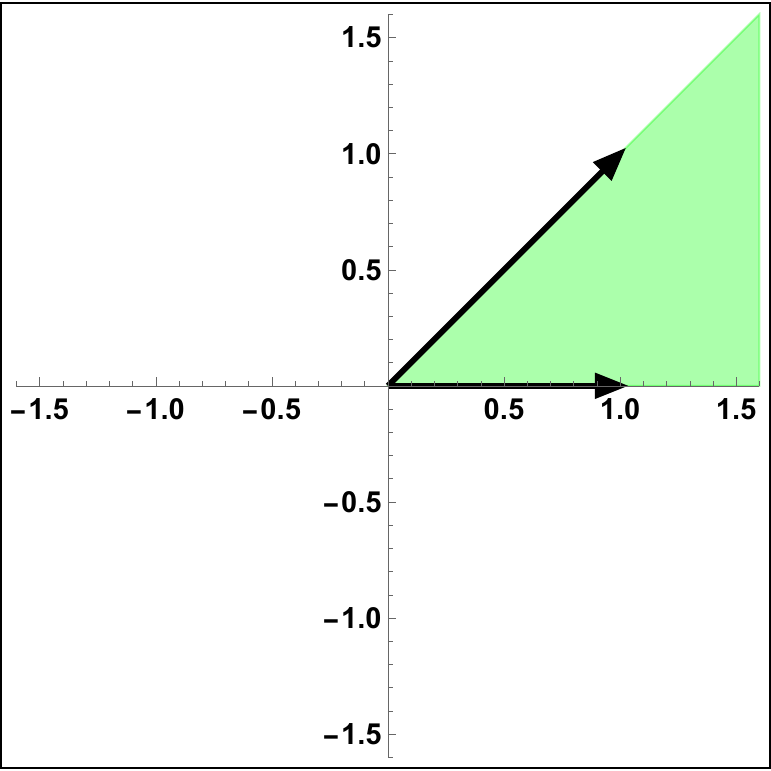}
       \caption{A two-dimensional conic hull.\label{Conic_Sample}}
   \end{figure}

But what is a conic hull? Conic hulls are semi-infinite geometric regions
 \begin{align}
      \{ p + s_1v_1 + \cdots + s_n v_n | s_i \in \mathbb{R}_{+} \}
  \end{align}
  where, the point $p$ is a vertex and the $v_i$'s are basis vectors. 
  For example, if $p=(0,0)$ and $v_1=(1,0),v_2=(1,1)$, then one gets the conic hull shown of Fig. \ref{Conic_Sample}.
   
  At Step 3 above, for a given $N$-fold MB integral, the possible conic hulls are constructed from the ${\bf e_i}$ (of the gamma functions in the retained $N$-combinations) which play the role of basis vectors.
   
   Let us now focus on Step 4. For complicated MB integrals, there can be hundreds or even thousands of different series representations, each of these being a linear combination of many multivariable series selected in a large set of such objects. For instance, the MB representation of the off-shell massless conformal hexagon, which is a 9-fold MB integral \cite{Loebbert:2019vcj, Ananthanarayan:2020ncn}, has in the non-resonant case 194160 different series representations, all built from linear combinations of dozens of series that belong to a set 2530 different series of 9 variables. Using other computational approaches of Feynman integrals, such as the negative dimension method \cite{Halliday:1987an, Dunne:1987am}, the Yangian approach \cite{Loebbert:2019vcj} or the method of bracket \cite{Gonzalez:2010zzb, Ananthanarayan:2021not}, one would also find this set of 2530 series, however, in order to find the possible series representations built from the latter (and in fact, even to find a single one...), one would need to know the convergence regions of each of the 2530 series and study their possible intersections\footnote{Note that this would also be the case with the MB method developed in \cite{Sasiela}.}, which obviously is a formidable, if not impossible task.
   
This is where the conic hulls method provides an important improvement compared to the alternative methods mentioned above: this very complicated convergence analysis is completely bypassed by the introduction of an appropriate set of conic hulls, as it is then sufficient to find their largest intersecting subsets in order to find the series representations of the MB integral. Moreover, as a bonus described in Step 5, each relevant intersection of conic hulls can be used to find a master conic hull, from which one can obtain the region of convergence (or a region included in the latter) of the series representation, by finding the master series' ones mapped back from it. This is a conjecture that has been checked in many examples. 

   
Let us note that, in its first version, the conic hulls method was developed for MB integrals with non-straight contours satisfying the usual property described in Section \ref{Section1}. After applying it with success to the computation of difficult non-resonant and resonant MB representations of Feynman integrals \cite{Ananthanarayan:2020ncn, Ananthanarayan:2020xpd}, we focused on the case of arbitrary straight contours of integration (parallel to the imaginary axis in the complex planes of the integration variables) that we solved in  \cite{Banik:2022bmk} and implemented in a new version of \texttt{MBConicHulls.wl}. The solution to this problem is simple: it is sufficient to use the generalized reflection formula 
  \begin{align}
  \Gamma(z-n)=\frac{\Gamma(z)\Gamma(1-z)(-1)^{n}}{\Gamma(n+1-z)}
  \end{align}
 in such a way that the real part of each gamma function in the numerator of the MB integrand becomes positive along the contours. One can then apply the conic hulls method in the same way as in the non-straight contour case.

\subsection{The triangulation method}
     
The conic hulls method is very efficient, but still too slow when dealing with (very) complicated MB integrals. Trying to improve the method, we found that regular triangulations of particular point configurations are dual to
the relevant intersections of conic hulls \cite{Banik:2023rrz} and, therefore, that they can be used to find the different series representations of a given MB integral. The great advantage of triangulations is that, once automatized, they provide the results for complicated integrals in a much faster way than conic hulls\footnote{The computational times of the two methods are compared in \cite{Banik:2023rrz}.}.
We have thus upgraded the \texttt{MBConicHulls.wl} package in such a way that now, in addition to the original conic hulls analysis, it offers the possibility to compute MB integrals from triangulations of point configurations. This is achieved by incorporating in the package the \texttt{TOPCOM} software \cite{Rambau:TOPCOM:2002} dedicated to triangulations. This way, more complicated MB integrals can be calculated analytically.

To apply the method, one needs to perform a change of the integration variables in order to rewrite Eq. \eqref{NfoldMB} in the \textit{canonical form}
\begin{equation} \label{N_MB_2}
    I_N= \int\limits_{-i \infty}^{+i \infty} \frac{ d z_1}{2 \pi i} \cdots \int\limits_{-i \infty}^{+i \infty}\frac{ d z_N}{2 \pi i}\,\,  \frac{ \Gamma(-z_1)\cdots\Gamma(-z_N) \prod\limits_{i=N+1}^{k'} \Gamma^{a'_i}(s'_i ({\bf z})) }{\prod\limits_{j=1}^{l} \Gamma^{b'_j}(t'_j ({\bf z}))} x'^{z_1}_{1} \cdots x'^{z_N}_{N}
\end{equation}
where we have pulled out the factors $ \Gamma(-z_1)\cdots\Gamma(-z_N) $ in the numerator. One can then build the needed configuration of points as a set of $N$ points
  \begin{equation}
P_1 = 
e'_{i 1} \,\, , \hspace{1cm}
P_2 = 
e'_{i 2} \,\, , \hspace{0.5cm} \cdots \hspace{0.5cm}
P_{N} = 
e'_{i N}
\end{equation}
and $\sum_{i=N+1}^{k'}a_i'$ additional points corresponding to the unit vectors of dimension $\sum_{i=N+1}^{k'}a_i'$. This set of $N+\sum_{i=N+1}^{k'}a_i'$ points is the one on which the triangulations are performed.

We applied the method on various Feynman integrals having MB representations with a high number of folds. Coming back to the hexagon and double-box \cite{Ananthanarayan:2020ncn}, for instance, we could show that they have respectively 194160 and 243186 different series representations and, exploring these large sets, that there are simpler series representations than those obtained in \cite{Ananthanarayan:2020ncn} in the non-resonant case and in \cite{Ananthanarayan:2020fhl} in the resonant case. We also used the off-shell one loop scalar massless $N$-point Feynman integral with generic powers of the propagators to test the computational possibilities of the triangulation approach. At $N=15$ (the MB representation of the 15-point integral has 104 folds) \texttt{TOPCOM} seems to reach its limits but it is still possible to obtain some triangulations with the help of an average personal computer.

\section{Conclusions}

We have shown that $N$-fold MB integrals can be analytically computed in a non-iterative way by introducing conic hulls or triangulations of point configurations. Both the resonant and non-resonant cases can be handled by these methods which are automatized in the \texttt{MBConicHulls.wl} package. This package was used to perform the first analytic computation of the massless off-shell conformal hexagon and double box for generic and unit powers of the propagators. It can also be used to study the transformation theory of multivariable hypergeometric functions, which has been largely unexplored, in a systematic way. 
However, a lot of work remains to be done for a full understanding of the field of investigations that these new approaches have opened. For instance, one would like to apply them to the calculation of MB representations of Feynman integrals having less scales than integration variables (presently, this can be done in certain cases but not all). 
It would also be interesting to better understand the master series conjecture and to find the rigorous link between triangulations and conic hulls. Another problem concerns the ``white zones" which are regions where none of the series representations extracted from a direct calculation of many MB integrals converge: finding a systematic way to obtain series representations valid in these regions would be important. At last, a deeper study of non-degenerate MB integrals  and their links with asymptotics in several variables, resurgence, etc. will probably yield interesting results.

\bigskip

\noindent {\bf Acknowledgements}

\medskip
S. F. thanks the organizers of MTTD23 for the very nice atmosphere of the conference and the interesting talks and discussions.

\end{document}